\documentclass[floats,amsmath,amssymb,pra,aps,showpacs,superscriptaddress,twocolumn]{revtex4-1}

\pdfoutput=1
\usepackage{amsmath,amsfonts,amssymb,amsthm,graphics,graphicx,epsfig,bbm}
\usepackage[colorlinks=true,citecolor=blue,linkcolor=blue,urlcolor=blue]{hyperref}
\usepackage[usenames]{color}
\usepackage{graphicx}
\usepackage{subfigure}
\usepackage{amsmath}
\usepackage{epsfig}
\usepackage{dcolumn}
\usepackage{bm}
\usepackage{color}
\usepackage{epstopdf}
\usepackage{amssymb}
\usepackage{amstext}
\usepackage{latexsym}
\usepackage{hyperref}
\usepackage{amsfonts}
\usepackage{psfrag}
\usepackage{xcolor}
\usepackage[normalem]{ulem}
\usepackage{dsfont}
\usepackage{txfonts}
\usepackage{ifthen}

\usepackage{subfigure,float,morefloats}

\graphicspath{{./Images/}{./AppendixImages/}}

\newcommand{\rev}[1]{#1}

\newcommand{\psicc}{\psi_{\text{c}}}
\newcommand{\psir}{\psi_{\text{ring}}}

\newcommand{\ket}[1]{\ensuremath{\left\vert #1 \right\rangle}}

\newcommand{\braket}[2]{\left\langle #1 \,\mskip1mu\vrule\mskip1mu\, #2 \right\rangle}

\newcommand{\an}[2]{\ensuremath{\hat{#1}^{\protect\phantom{\dagger}}_{#2}}}
\newcommand{\cn}[2]{\ensuremath{\hat{#1}^\dagger_{#2}}}

\newcommand{\expU}[1]{\ensuremath{e^{#1}}}
\newcommand{\abs}[1]{\left|#1\right|}

\newcommand{\nnl}{\ensuremath{\beta}}

\newcommand{\eexp}[1]{\mathrm{e}^{#1}}

\newcommand{\be}[1]{\begin{eqnarray} \label{e#1}}
\newcommand{\beq}{\begin{eqnarray}}
\newcommand{\eeq}{\end{eqnarray}} 
\newcommand{\hide}[1]{}
\newcommand{\Eq}[1]{{Eq.\!\!~(\ref{#1})}} 

\newcommand{\hrefl}[1]{\href{#1}{[link]}}

\newcommand{\vc}[1]{\ensuremath{{\bf#1}}}

\newcommand{\op}[1]{\ensuremath{\hat{#1}}}
\newcommand{\avg}[1]{\ensuremath{\langle#1\rangle}}

\newcommand{\subfigimg}[3][,]{%
		\setbox1=\hbox{\includegraphics[#1]{#3}}
		\leavevmode\rlap{\usebox1}
		\rlap{\hspace*{2pt}\raisebox{\dimexpr\ht1-0.5\baselineskip}{{\bfseries \large\textsf{#2}}}}
		\phantom{\usebox1}
}

\newcommand{\idg}[1]{{\bfseries #1)}}

\begin{document}

\author{Tobias Haug}
\affiliation{Centre for Quantum Technologies, National University of Singapore,
	3 Science Drive 2, Singapore 117543, Singapore}
\author{Joel Tan }
\affiliation{Centre for Quantum Technologies, National University of Singapore,
	3 Science Drive 2, Singapore 117543, Singapore}
\author{Mark Theng}
\affiliation{Centre for Quantum Technologies, National University of Singapore,
	3 Science Drive 2, Singapore 117543, Singapore}
\author{Rainer Dumke}
\affiliation{Centre for Quantum Technologies, National University of Singapore, 3 Science Drive 2, Singapore 117543, Singapore}
\affiliation{Division of Physics and Applied Physics, Nanyang Technological University, 21 Nanyang Link, Singapore 637371, Singapore}
\author{Leong-Chuan Kwek}
\affiliation{Centre for Quantum Technologies, National University of Singapore,
	3 Science Drive 2, Singapore 117543, Singapore}
\affiliation{MajuLab, CNRS-UNS-NUS-NTU International Joint Research Unit, UMI 3654, Singapore}
\affiliation{Institute of Advanced Studies, Nanyang Technological University,
	60 Nanyang View, Singapore 639673, Singapore}
\affiliation{National Institute of Education, Nanyang Technological University,
	1 Nanyang Walk, Singapore 637616, Singapore}
	\author{Luigi Amico}
\affiliation{Centre for Quantum Technologies, National University of Singapore,
	3 Science Drive 2, Singapore 117543, Singapore}
\affiliation{MajuLab, CNRS-UNS-NUS-NTU International Joint Research Unit, UMI 3654, Singapore}
\affiliation{Dipartimento di Fisica e Astronomia, Via S. Sofia 64, 95127 Catania, Italy}
\affiliation{CNR-MATIS-IMM \&   INFN-Sezione di Catania, Via S. Sofia 64, 95127 Catania, Italy}
\affiliation{LANEF {\it 'Chaire d'excellence'}, Universit\`e Grenoble-Alpes \& CNRS, F-38000 Grenoble, France}

\title{Readout of the atomtronic quantum interference device}

\begin{abstract}
A Bose-Einstein condensate confined in ring shaped lattices interrupted by a weak link and pierced by an effective magnetic flux defines the atomic counterpart of the superconducting quantum interference device: the atomtronic quantum interference device (AQUID). In this paper, we report on the detection of current states in the system through a self-heterodyne protocol. 
Following the original proposal of the NIST and Paris groups, the ring-condensate many-body wave function interferes with a reference condensate expanding from the center of the ring. We focus on the rf-AQUID which realizes effective qubit dynamics. Both the Bose-Hubbard and Gross-Pitaevskii dynamics are studied. For the Bose-Hubbard dynamics, we demonstrate that the self-heterodyne protocol can be applied, but higher-order correlations in the evolution of the interfering condensates are measured to readout of the current states of the system. We study how states with macroscopic quantum coherence can be told apart analyzing the noise in the time of flight of the ring condensate.
\end{abstract}
\maketitle

\section{Introduction}
Atomtronics exploits the progress in quantum technology 
to realize  atomic circuits   in which  ultra-cold atoms are manipulated in versatile  laser generated or magnetic guides\cite{Rainer_Birkl,BIRKL200167,Schlosser2011,Boshier_painting,keil2016fifteen,Amico_Atomtronics,Amico_NJP}.    Although atomtronic   circuits 
quantum devices and simulators may be of a radically different type  from current technology, a fruitful starting point in the current  research has been considering ultracold matter-wave-analog of known quantum electronic systems.
With this logic, ring-shaped  condensates  interrupted  by  one or several weak  links and pierced by an effective magnetic flux\cite{dalibard2011colloquium},  have been studied  in analogy with the SQUIDs of  mesoscopic
superconductivity \cite{wright2013driving,Ramanathan2011,Ryu2013}.  Such systems, dubbed  Atomtronics  Quantum
Interference Devices (AQUIDs), with enhanced control of noise and low decoherence, enclose a great potential both for basic science and technology.
In particular,  for AQUIDs with  weak barriers and  weak atom-atom interaction, hysteresis effects were evidenced\cite{eckel2014hysteresis}. In this case, the system can serve to study the dynamics of vortices in a quantum fluid with a new twist\cite{yakimenko2015}. Such study may give  important contributions to long-standing problems in quantum turbulence\cite{paoletti2011quantum}. 
For stronger interactions and higher barriers,  AQUIDs were demonstrated to be governed by an effective two-level system (qubit) dynamics\cite{hallwood2006macroscopic,solenov2010metastable,amico2014superfluid,aghamalyan2015coherent,aghamalyan2016atomtronic,Mathey_Mathey2016}.

In this paper, we consider  AQUIDs in which  a single weak link is present (that, in analogy with quantum electronic devices, defines the rf-AQUID); additionally, a  lattice potential along the azimuthal angle is applied\cite{amico2014superfluid, aghamalyan2015coherent}. 
The resulting device  can be indeed considered as   the  cold  atoms  analogue  of  the  many-Josephson-junction  fluxonium\cite{manucharyan2009fluxonium}.

Depending on the  conditions of the system (atomic density, atom-atom interaction, external effective magnetic flux, strength of the weak-link), the flow of atoms in AQUIDs  entails physical  mechanisms that may be very different in nature. In the simplest situation, the current is made of  atoms  in a definite (azimuthal) angular momentum state. By  tuning the effective magnetic field suitably, it is however possible to put {\it each} particle in the ring in a superposition state of two angular momentum states. Finally, in the qubit dynamics regime,  the current states are  indeed many-particles entangled states with macroscopic quantum coherence made of  symmetric and anti-symmetric combinations of the clockwise and anti-clockwise flowing-states (cat states)\cite{leggett1980macroscopic}.  

Although the existence of the atomic current flowing  in AQUID can be detected by standard time-of-flight measurement of the ring condensate\cite{aghamalyan2015coherent}, the analysis  of the specific features of the flowing states needs  a more sophisticated  configuration. Specifically, the ring  condensate is allowed to interfere with a second condensate confined with Gaussian-shaped laser beam in the center of ring. Such condensate sets the reference for the phase winding of the ring condensate. By {\it in-situ} measurement of the two interfering condensates,  such protocol provides, indeed, the self-heterodyne detection of the phase of the wave function of the particles trapped in  the ring potential.  
With such an approach, it  was demonstrated that  both  the orientation and the intensity of the current states can be detected\cite{eckel2014interferometric,corman2014quench,mathew2015self} (see also\cite{roscilde2016quantum}). Very recently, it was suggested that similar information can be obtained by analyzing the Doppler shift of the phonon modes caused by shrinking the flowing condensate\cite{StringariDoppler2016}. 

In such analysis the ring condensate was assumed to be in the hydrodynamic Gross-Pitaevskii regime (weak interaction). 

The read-out of the current states in AQUID is very important to be carried out for quantum simulation. At the same time,  it is a crucial step to fill for any application of the AQUID for quantum computing.
In this paper, we apply the self-heterodyne  approach to read-out the current states in rf-AQUIDs made of ring-lattice of condensates interrupted by a single weak-link. We study the systems both in Gross-Pitaevskii and Bose-Hubbard regimes. In the latter one, appropriate for stronger atom-atom  interactions, we demonstrate that the standard measurement of the time evolution of the statistical average of  particle density does not reveal the main properties of the current states  (orientation  and intensity of the flow). The read out of the current states, instead,  can be done  by considering the  averaged density-density correlations in two different positions of the condensates. At the same time, our analysis provides a way to detect  states with macroscopic quantum coherence made of superpositions of  clockwise and and anti-clockwise flows. 

The article is structured as follows. In Sect. \ref{physical_model}, we summarize the main physical properties of the rf-AQUID together with the models we exploit to describe the system.
In Sect.\ref{int_phase}, we describe the protocol leading to the heterodyne detection of the phase in ring condensates. Similarities and difference between the Gross-Pitaevskii and Bose-Hubbard dynamics are discussed. In Sect.\ref{results}, we present the results we achieve. Discussions, comments and remarks are presented in the closing  Sect.\ref{conclusions}. In the appendix, we provide supplementary results on the Bose-Hubbard and Gross-Pitaevskii dynamics, and provide the analytical calculations for the noninteracting system we used to benchmark our results.


\section{The Atomtronic quantum interference device}
\label{physical_model}

The relevant physics of the system is captured by the Bose-Hubbard Model. The Hamiltonian reads
\begin{equation}
\mathcal{H}_{\text{BH}}  =  \sum_{i=1}^{M} \left[
\frac{U}{2} \bm{n}_{i}(\bm{n}_{i}-1)+\Lambda_i\bm{n}_{i}
- {J_i} \left(\eexp{-i2\pi\Omega/M} \bm{a}_{i{+}1}^{\dag} \bm{a}_{i} + \text{h.c.} \right)
\right]
\label{BH}
\end{equation}
where $\bm{a}_i \, (\bm{a}^{\dagger}_i)$ are bosonic annihilation (creation) operators on the $i$th site of a ring with length $M$
and $n_{i}= \bm{a}_{i}^{\dag}\bm{a}_{i}$ is the corresponding number operator.
Periodic boundaries are imposed, meaning that $\bm{a}_{M} \equiv \bm{a}_0$.
The parameter $U$ takes into account the finite scattering length for the atomic 
two-body collisions on the same site: $U=4 \pi \hbar^2 a_0\int dx |w(x)|^4/m$,  $w(x)$ being the Wannier functions of the lattice, $m$ the mass of atoms and $a_0$ the scattering length. To break the translational symmetry, there are two possible ways: Either, the  hopping parameters are all equal ${J_i=J}$ 
except in  one weak-link hopping $i_0$ where  ${J_{i_0}=J'}$. \rev{The other alternative, which we choose in this paper, is to place a potential barrier at a single site ${\Lambda_i=\Lambda }$ and at all other sites the potential is set to zero, with $J_i=J,  \forall i$. The two options show qualitatively the same physics\cite{aghamalyan2015coherent}}.
The ring is pierced by an artificial (dimensionless) magnetic flux $\Omega$,
which can be experimentally induced for neutral atoms as a Coriolis flux by rotating the lattice
at constant velocity~\cite{fetter,wright2013driving}, 
or as a synthetic gauge flux by imparting a geometric phase
directly to the atoms via suitably designed laser fields~\cite{berry,synth1,dalibard2011colloquium}.
The presence of the flux $\Omega$ in \Eq{BH} has been taken into account through 
the Peierls substitution: $J_i \rightarrow e^{-i2\pi\Omega/M} J_i$. 
{The Hamiltonian~(\ref{BH}) is manifestly periodic in $\Omega$ with period $1$.
In the absence of the weak-link, the system is also rotationally invariant and therefore the particle-particle interaction energy does not depend on $\Omega$. 
The many-body ground-state energy, as a function of $\Omega$, is therefore given by a set of parabolas 
intersecting at the frustration points ${\Omega_{n} = (n + \frac{1}{2})}$~\cite{leggett,loss}. 
The presence of the weak-link breaks the axial rotational symmetry and couples different angular momenta states, thus lifting the degeneracy at $\Omega_{n}$ (see the Appendix \ref{appendix_perurbation}). This feature sets the qubit operating point\cite{amico2014superfluid,aghamalyan2015coherent}.}  

It is worth noting that  the interaction $U$ and the weak-link strength induce competing physical effects: the weak-link sets an  healing  length in the density as a 
further spatial scale; the interaction  tends to smooth out the healing length effect. As a result, strong interaction tends to renormalize the weak link 
energy scale\cite{aghamalyan2015coherent,cominotti2014optimal}.

Assuming that the quantum dynamics of the system can be described by the quantum dynamics of the phase of the bosons $a_i\sim \sqrt{ \bar{n}}e^{i\phi_i} $,  the  effective dynamics 
of the AQUID is governed by \cite{amico2014superfluid}
%
\begin{equation}
\mathcal{H}_{\text{eff}} \ \ = 
 \mathcal{H}_{\text{syst}}+
 \mathcal{H}_{\text{bath}} +  \mathcal{H}_{\text{syst-bath}}
\label{H_eff}
\end{equation}
with 
\begin{equation}
\mathcal{H}_{\text{syst}}=U  \bm{n}^2 
+ \frac{1}{2}E_L \bm{\varphi}^2
- E_J \cos(\bm{\varphi}-2\pi\Omega) \;,
\label{jj}
\end{equation} 
in which $ \varphi$ is the phase slip across the weak link, $E_L=J/M$, and $E_J=J'$.  
For $\delta \doteq {E_J}/{E_L} \ge 1 $, $\mathcal{H}_{\text{syst}}$ describes a particle in a double well potential with the two-minima well separated from the other
 features of the potential.  The two parameters, U and $J'/J$, allow control of the two level system. The  two local minima of the double well are degenerate for  ${\Omega=\frac{1}{2}}$.
The minima correspond to the clock-wise and anti-clockwise currents in the AQUID.

\rev{In the  weak interaction regime and coherence length in the condensate much larger than the lattice spacing, the many-body wave function eigen-solution of  Eq.(\ref{BH}) can be considered as a product of coherent single particle wave functions (such conditions could be met for sufficiently shallow lattice).} This is the  limit in which the  system's dynamics can be simplified to obey the  Gross-Pitaevskii equation.  Incidentally, we observe that, by construction, the Gross-Pitaevskii  Bose fluid cannot contain  $N$ particle entanglement (as e.g. cat states entanglement). The kind of coherence possibly encoded in the system in such regime, instead, may arise by superposition states of single particles.  The Gross-Pitaevskii equation reads

\begin{equation}
i \hbar \partial_t \psi(\vc{r}) = \left( - \frac{\hbar^2}{2m}\nabla^2 + V(\vc{r}) + g \left|\psi(\vc{r})\right|^2 - \Omega\Omega_0 L_z\right)\psi(\vc{r})\;,
\end{equation}
where $ \psi(\vc{r})$ is the so called condensate wave function.
The function $V(\vc{r})$ denotes the external trap potential, $N$ is the total number of atoms, $ g = \frac{4 \pi \hbar^2 a_0 N}{m} $ is the coupling constant, $ a_0 $ is the scattering length, $N$ the number of atoms,
$ \Omega $ is the effective flux imparted by the rotation of the weak link, $ L_z = -i \hbar \left(x \partial_y - y \partial_x\right) $ is the angular momentum operator, and $ \Omega_0 = \frac{\hbar}{m R^2} $ is the characteristic frequency and flux quantum of the ring.





We model the trap with a ring Gaussian potential. Modulating the potential with an azimuthal envelope allows us to add a lattice and a weak link. 
We assume a strong confinement of the condensate in $z$ direction, restricting the non-trivial dynamics to two dimensions in the $x-y$ plane.
The full expression for the potential in polar coordinates is
\begin{equation}\label{potential_equation}
V(r,\phi) = V_0V_\text{ring}(r)\left(1 + \eta_\text{l}\sin^2 \left(\frac{M}{2}\phi\right) - \eta_\text{w} e^{-\frac{\phi^2}{\xi_\text{w}^2}}\right)\;,
\end{equation}
where $r$ and $\phi$ are the radial and azimuthal coordinates, $V_0$ is the strength of the trap and ${V_\text{ring}(r)=-e^{-\frac{\left(r - R\right)^2}{\xi_\text{r}^2}}}$ is a ring Gaussian potential with ring radius $R$ and ring potential width $\xi_\text{r}$. The second term in the brackets adds a lattice with $n$ sites and relative strength $\eta_\text{l}$. The third term adds a weak link of relative strength $\eta_\text{w}$ and angular size $\xi_\text{w}$. 


\rev{We solve the normalized 2-dimensional GPE equation, and parametrize the interaction nonlinearity with a dimensionless parameter $\beta$\cite{bao2003numerical}. The scattering length is then ${a_0=\frac{\beta \sigma_\text{z}}{2\sqrt{2\pi}N}}$, where $\sigma_\text{z}$ is the characteristic length of a harmonic confinement in $z$ direction. For ${}^{87}\text{Rb}$ atoms (${a_0\approx 100a_\text{Bohr}}$), ring radius ${R=10\mu\text{m}}$, ${\sigma_\text{z}=0.2R}$, ${N=10^{3}}$, radial confinement ${\sigma_\text{r}=0.083R}$ we find ${\beta=13.3}$. The corresponding density of atoms in the ring is approximately ${n=\frac{N}{2\pi 2\sigma_\text{r}2\sigma_\text{z}R}=2.4\cdot10^{12}\text{cm}^{-3}}$.

To go the Bose-Hubbard regime, the lattice depth is increased such that each lattice site is well localized. Then, the following approximations for the Bose-Hubbard parameters can be used: ${J=\frac{4}{\sqrt{\pi}}E_\text{r}s^{3/4}\exp(-2s^{1/2})}$ and ${U=\frac{8}{\sqrt{\pi}}k_\text{L}a_0 E_\text{r} s^{3/4}}$,
with the recoil energy $E_\text{r}=\frac{\hbar^2k_\text{L}^2}{2m}$, the lattice depth $V_0$, ${k_\text{L}=\frac{\pi}{a_\text{lattice}}}$, the lattice constant $a_\text{lattice}$ and the ratio ${s=\frac{V_0}{E_\text{r}}}$\cite{zwerger2003mott}. For Rubidium atoms, ${R=10\mu\text{m}}$, number of lattice sites ${M=14}$ and ${s=10}$, we find ${J/\hbar=4.1\,\text{Hz}}$ and ${U/J=1.46}$.}

\section{Interferometric detection of the current states}
\label{int_phase}
As discussed in the previous section, 
the atomic current  is provided by   an imparted phase gradient of the wavefunction along the ring condensate.  
To read-out the direction and the intensity of the current in our lattice system, we follow the logic originally applied in a series of works carried out by the Maryland and Paris groups to map-out the circulating states in continuous ring-shaped condensates\cite{moulder2012quantized,wright2013driving,eckel2014hysteresis,eckel2014interferometric,corman2014quench}.  Accordingly, the ring condensate is made to  interfere with a Gaussian condensate at rest, located at the center, fixing the  reference for the phase of the wavefunction. 
The combined wavefunction evolves in time, interferes with itself and finally is measured.  The number of spirals gives the total number of rotation quanta.

In the actual experiment,  the condensate is imaged through in-situ measurements. In this way, the current direction and magnitude is well visible as a spiral pattern.
The position of the spirals depends on the relative phase between ring and the central condensate ${\psi=\psir+\expU{i\phi}\psicc}$. 

In the theoretical explanations provided so far, the mean-field Gross-Pitaevskii equation was applied, which assumes that the combined system is a simple product of  one particle problems in a {\it coherent state} and in a 
 superposition of particles  being in the ring and in the central condensate.   In such a state,
the particle number is not conserved, and thus according to the particle number - phase uncertainty, the phase is well defined. The relative phase is simply a free parameter, chosen at random (the randomness being generated by experimental imperfections during the preparation).

However, these assumptions are not generically fulfilled. In the experiment,  the ring and central condensate are, indeed,  well separated for an extended period of time (thus the coherence between the two is lost) and  each of them have conserved particle numbers. They can be  described as product states of two Fock states ${\ket{\Psi}=\ket{\Psi_\text{r}}\otimes\ket{\Psi_\text{c}}}$. 
Thus, there is no a priori defined phase between the two.  A definite phase can arise when the information about the particle number is erased. Indeed, even if the two condensates do not have a phase relation, a distinct, random phase results when the position of most of the particles are measured\cite{castin1997relative,mullin2006origin}. In this measurement procedure, the information from which condensate the particles came from is erased. This implies that the relative phase is not a property of the two condensates (or a local hidden parameter), but is determined only when the measurement is performed. 


In the following, we consider two separate cases in which the ring lattice is  governed by Gross-Pitaevskii or Bose-Hubbard models.

\subsection{Bose-Hubbard dynamics}
\label{BHdynamics}

The ring wavefunction is calculated by solving the ground state of the Bose-Hubbard Hamiltonian, while the central condensate is simply a single decoupled site with $N_\text{c}$ particles. 
\rev{In a single experimental run, the spirals will be visible for a good condensate with high number of particles. 
 From a theoretical point of view, such single shot results  could be generated by obtaining the many-body eigenfunctions of Eq. (\ref{BH}), combining them with the wave function of the central condensate and simulating the detection sequence of all particles of the expanding wavefunction\cite{castin1997relative,mullin2006origin}. In our case, however, the latter approach does not  produce  the spiral patterns because the numerics are limited to a small number of particles. To overcome this limitation, we resort to expectation values, which experimentally corresponds to take averaged results over  many experimental runs.
Such an approach could be particularly helpful for systems in which the visibility in single shot experiments is low.}
The dynamics of the density $\op{n}(\vc{r},t)=\op{\psi}^\dagger (\vc{r},t)\op{\psi}(\vc{r},t)$ is initialized assuming that the bosonic field operator of the system is ${\op{\psi}(\vc{r})=\sum_n w_n(\vc{r})\an{a}{n}}$, where  $w_n(\vc{r})$ are a set of Wannier functions forming a complete basis\cite{altman2004probing,gerbier2008expansion}. In our calculation, we approximate the full basis for wave functions living in the ambient space on which the condensate expands with the set of  Wannier functions  composed of Gaussians peaked at the ring lattice sites and at its  centre  \rev{(the Gaussian approximation for the Wannier functions is a well verified approximation for single site wavefunctions -- see f.i. \cite{slater1952soluble,chiofalo2000collective})}. For the free evolution (we are indeed in a dilute limit) we  assume that each particle at site $n$ expands in two dimensions as  
\begin{equation}
w_n(\vc{r},t) = \frac{1}{\sqrt{\pi}}\frac{\sigma_n}{\sigma_n^2+\frac{i\hbar t}{m}}\expU{-\frac{(\vc{r}-\vc{r}_n)^2}{2\left(\sigma_n^2+\frac{i\hbar t}{m}\right)}} \;,\label{Gaussian}
\end{equation}
where $\sigma_n$ is the width of the condensate located at the $n$-th site. The dynamics of the condensates is then approximated as 
$\op{\psi}(\vc{r},t)=\sum_n w_n(\vc{r},t)\an{a}{n}$.
We observe that such approximation works well in the situations in which the optical lattice is assumed to be sufficiently dense in the space in which the condensate is released (as in the release from large three dimensional optical lattices). In our case, we checked that the scheme works for large ring lattices, and, at intermediate size, in the large number of particles  limit (see Appendix \ref{AppDens}). 
In particular, we checked that  the long time limit of the approximated density $\langle {\op{n}(\vc{r},t)\rangle =\langle \op{\psi}^\dagger(\vc{r},t)\op{\psi}(\vc{r},t)} \rangle$ coincides with the  time-of-flight expansion calculated by Fourier transforming the initial wavefunction.  


\rev{Generically, it is assumed that a single shot experiment with many particles (self-average), and the average over many realization would yield the same result. Here, however, this is not the case as every realization of the experiment has a random phase. This phase results in an interference pattern, which is averaged out over many repetitions.}
Indeed, we find that the density operator alone does not show any interference effects. 
This is consistent with the reasoning  given at the beginning of the section: As the relative phase between ring and central condensate is determined  randomly upon measurement, the expectation value of the density operator will average over different realizations of the spiral interference pattern, washing out the information on the current configuration structure. 
As the ring and central condensate are uncoupled, there is no uncertainty on whether a measured particle came from the ring or the central condensate. Accordingly, the density operator,  measuring  a  single particle property only, cannot give information on the interference between two condensates.  This is confirmed by Fig.\ref{densUtime} in the Appendix. However, if we measure two or more particles, information about the particle origin is lost\cite{castin1997relative,mullin2006origin}, as the measured particles could be either from the ring or from the central condensate. As the particle number distribution between ring and central condensate becomes uncertain, phase certainty is gained.

As we shall demonstrate below, we, indeed, observe an interference pattern in higher order density-density correlations. 
We calculate the density-density covariance\cite{castin1997relative,mullin2006origin,folling2005spatial,kang2014revealing}
\begin{equation}
\text{cov}(\vc{r},\vc{r'},t)=\avg{\op{n}(\vc{r},t)\op{n}(\vc{r'},t)}-\avg{\op{n}(\vc{r},t)}\avg{\op{n}(\vc{r'},t)}\;.
\end{equation}
We also define the root of the density covariance which has the same unit as the density to improve the contrast of the measured interference pattern
\begin{equation}
\sigma(\vc{r},\vc{r'},t)=\text{sgn}(\text{cov}(\vc{r},\vc{r'},t))\sqrt{\abs{\text{cov}(\vc{r},\vc{r'},t)}}\;.
\end{equation}

\subsection{Gross-Pitaevskii dynamics}
In the mean-field Gross-Pitaevskii description, the relative phase of two well separated condensates (e.g. ring and central condensate ${\psi=\psir+\expU{i\phi}\psicc}$) is a free parameter, in contrast to the Bose-Hubbard model. This phase factor shifts the spirals in azimuthal direction in the interferometric expansion. In experiment, this phase factor is determined randomly for each realization of the condensate, and thus will average out the spirals over many experimental runs. In the following, we consider the result of a single realization with a specific value of $\phi$.   

The ground state of the condensate in the ring potential is found  by imaginary time evolution of the Gross-Pitaevskii equation. After that, the Gaussian central cloud is added, the potential turned off and both ring and central condensate freely expand.

From \cite{mathew2015self}, an approximate formula for the expanding ring condensate (for zero interaction while expanding) is given by
\begin{equation}\label{green function analytics}
\psir(r,\theta,t) =\left(\expU{-\frac{(r-R)^2}{2\sigma_\text{r}(t)^2}}\varphi_\text{r}(\theta)+\expU{-\frac{(r+R)^2}{2\sigma_\text{r}(t)^2}}\varphi_\text{r}(\theta+\pi)\right)/\left(\mathcal{N}(t)\sqrt{r}\right) \;,
\end{equation}
with $\theta$ the angle in polar coordinates, $\mathcal{N}(t)$ a normalization factor, ${\varphi_\text{r}(\theta)}$ the initial angular wavefunction and ${\sigma_\text{r}(t)^2=\sigma_{\text{r}}^2+\frac{i\hbar t}{m}}$. Here, $ \sigma_{\text{r}} $ is the initial width of the radial profile of the wavefunction. Equation \ref{green function analytics} is valid for timescales ${\tau_\text{B}\ll t \ll \tau_\text{S}(r)}$, where ${\tau_\text{B}=\frac{m \sigma_{\text{r}}^2}{\hbar}}$ and ${\tau_\text{S}(r)=\frac{m r R}{\hbar}}$. $\tau_\text{B}$ is the timescale of the initial ballistic expansion of the ring cloud, while $\tau_\text{S}(r)$ is the timescale when the cloud acquires its far-field behaviour (it begins to look like its initial momentum distribution).

When ${ t < \tau_\text{B} \frac{r + R}{\sigma_{\text{r}}} }$, the second term is negligible. Interference with the expanding central cloud results in spiral fringes.
We define the characteristic time when fringes appear ${\tau=mR\sigma_r/\hbar}$. The shape of the fringes thus allows us to read out the phase profile of the initial state. 
As it is discussed below, the weak link induces a discontinuity in the phase of the initial state, resulting in the appearance of phase slips which cause a discontinuity in the spirals.

When $ t > \tau_\text{B} \frac{r + R}{\sigma_{\text{r}} }$, the second term becomes significant and interferes with the first term, resulting in the appearance of additional circular bands. Eventually, as ${ t \rightarrow \infty }$, the condensate evolves towards the Fourier transform of the initial state, which corresponds to the initial momentum distribution\cite{bloch2008many}.



Consistently with this theorem, we found that for sufficiently weak interactions and large time scales, the condensate at long times matches the momentum profile of the original wavefunction. 
We find that when the central cloud is co-expanded with the ring condensates, then the shape of the combined condensates at intermediate time scales shows a characteristic spiral pattern which reveals the phase winding.

\section{Results}
\label{results}

Our first goal will be pointing out the difference between the Gross-Pitaevskii and Bose-Hubbard dynamics.
Then, we will discuss the specific spiral pattern arising when the effective magnetic field is fixed to the degeneracy point ${\Omega=\frac{1}{2}}$. Previous studies suggest that the spiral pattern displays a specific  discontinuity at the position of the weak link. Such a discontinuity, however may reflect different states in terms of their entanglement.

\subsection{Bose-Hubbard interferometric measurement}
\begin{figure*}[htbp]
	\centering

	\includegraphics[width=0.9\textwidth]{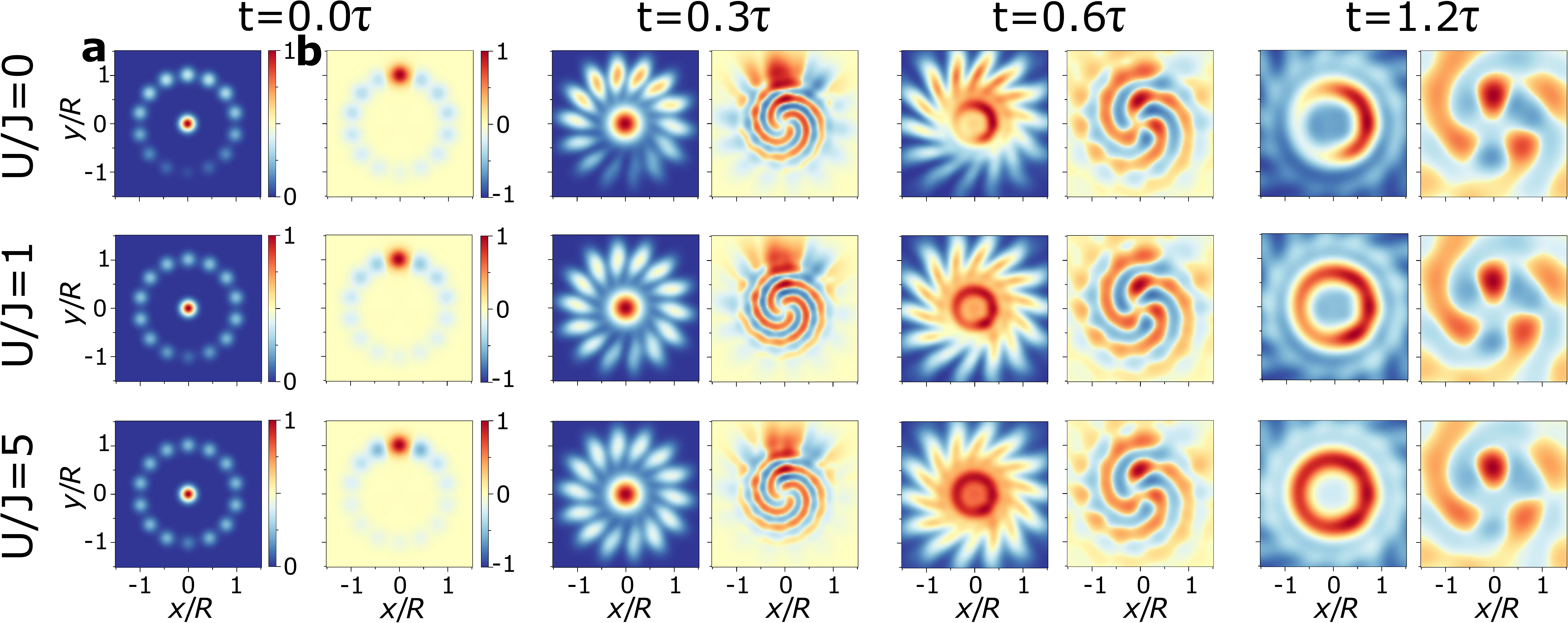}
	\caption{Density distribution \idg{a} $\avg{\op{n}(\vc{r})}$ and root of density-density covariance \idg{b} ${\sigma(\vc{r},{\vc{r'}=\{0,R/2\}})}$ of expanding atoms at times ${t=0,\,0.3\tau,\,0.6\tau,\,1.2\tau}$, with ${\tau=mR\sigma_r/\hbar}$. Calculated using Bose-Hubbard model, no interaction during expansion. Data in color and normalized to one. From top to bottom: ${U=0}$, ${U/J=1}$, ${U/J=5}$. 7 particles, ${M=14}$ ring sites, ring radius $R$. Width of central and ring cloud is ${\sigma_r=2 R/L}$ and potential barrier ${\Lambda=J}$, ${\Omega=3}$, 25\% of atoms in central condensate. Barrier at ${x=0}$, ${y=-R}$. 
	At intermediate times, we observe some spiral-like structure at the edges. This is not due to interference with the central condensate, but a residue of the ring lattice condensate interfering with itself. }
	\label{genFlux}
\end{figure*}

\begin{figure}[htbp]
	\centering

	\includegraphics[width=0.45\textwidth]{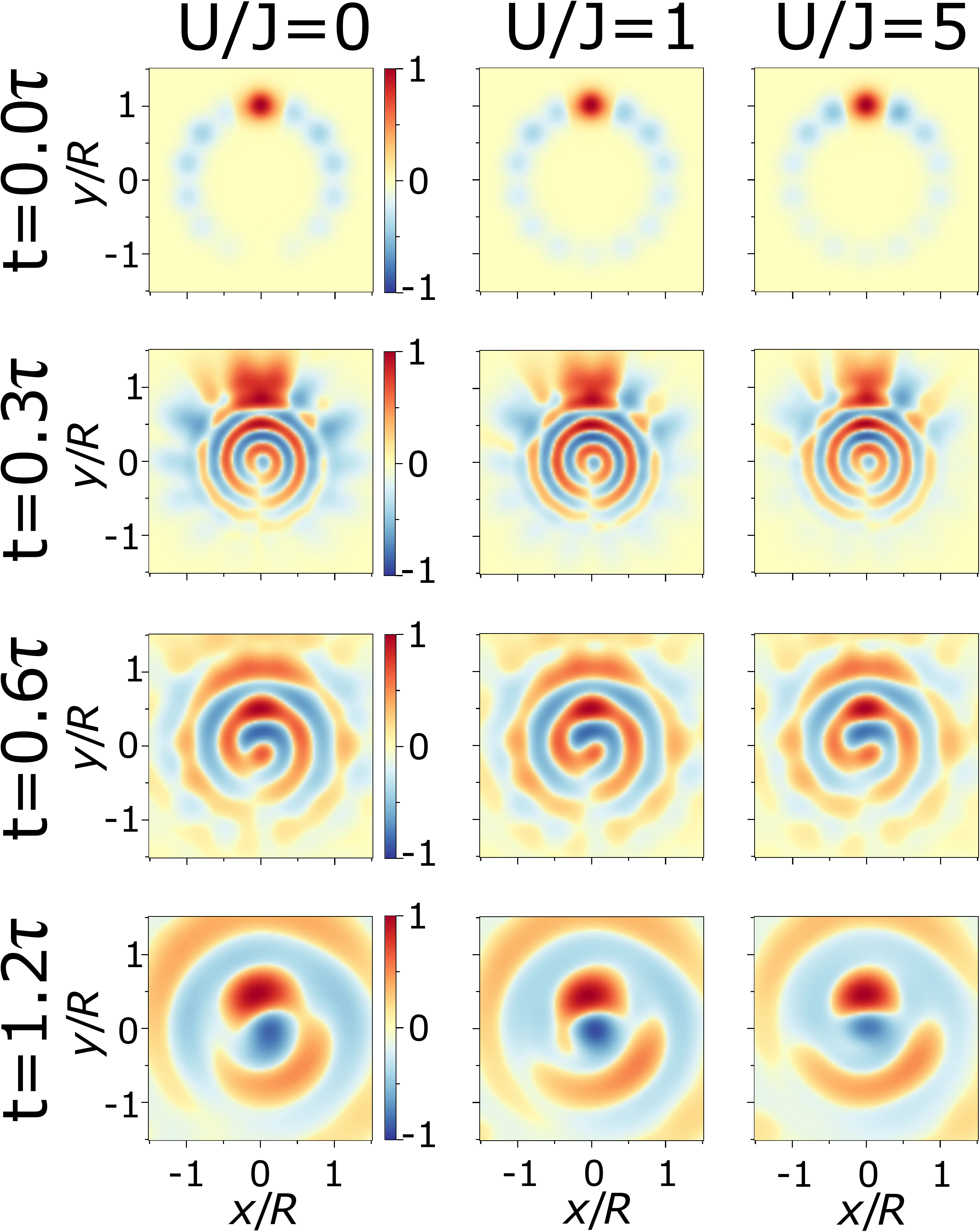}
	\caption{Root of density-density covariance ${\sigma(\vc{r},{\vc{r'}=\{0,R/2\}})}$ of expanding atoms at times ${t=0,\,0.3\tau,\,0.6\tau,\,1.2\tau}$, with ${\tau=mR\sigma_r/\hbar}$. Calculated using Bose-Hubbard model. From left to right: ${U=0}$, ${U/J=1}$, ${U/J=5}$. Flux ${\Omega=\frac{1}{2}}$ at the degeneracy point, other parameter same as in Fig.\ref{genFlux}. The discontinuity in the bottom of the spirals at intermediate times $t=0.3\tau$ and $t=0.6\tau$ shows that the ring condensate is in a superposition of zero and one rotation quantum. }
	\label{corrUtime}
\end{figure}
The first  conclusion comes from the analysis of the time evolution of the density  expectation value after the interference with the central condensate: Consistently with the reasoning reported above, we observe that no spiral appears in the density in Fig.\ref{genFlux}a.  In contrast, a clear spiral pattern arises in the  density-density covariance $\sigma(\vc{r},\vc{r'})$ in Fig.\ref{genFlux}b. \rev{We choose ${\vc{r'}=\{x'=0,y'=R/2\}}$ such that it maximizes the spiral visibility.}

Now, we present the results at the degeneracy point  ${\Omega=\frac{1}{2}}$--Fig.\ref{corrUtime}. In this case,  a step in the spirals at the weak link  site (here at the center bottom) is clearly visible for intermediate times. 

%

\begin{figure}[htbp]
	\centering
	\subfigimg[width=0.23\textwidth]{a}{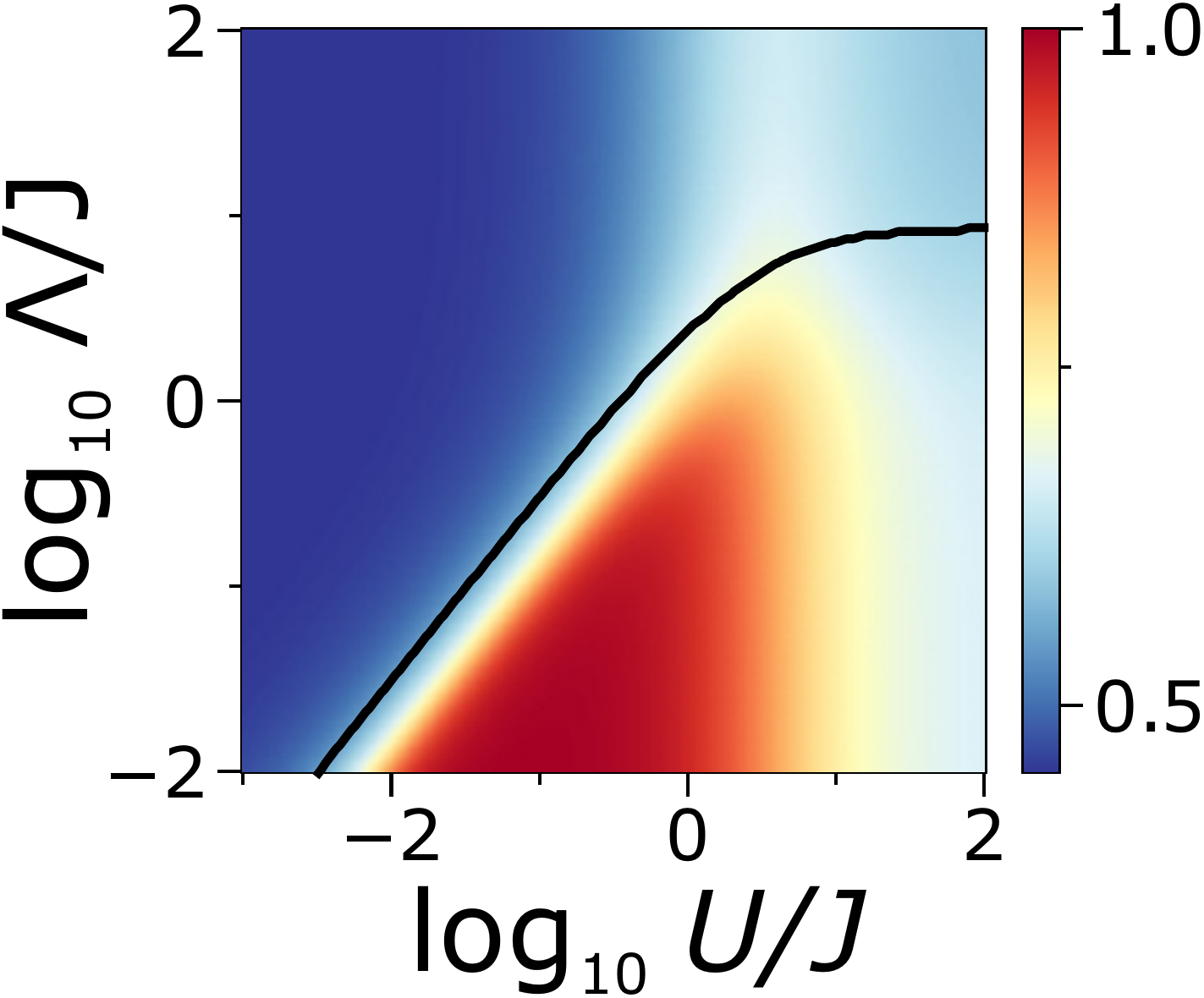}
	\subfigimg[width=0.23\textwidth]{b}{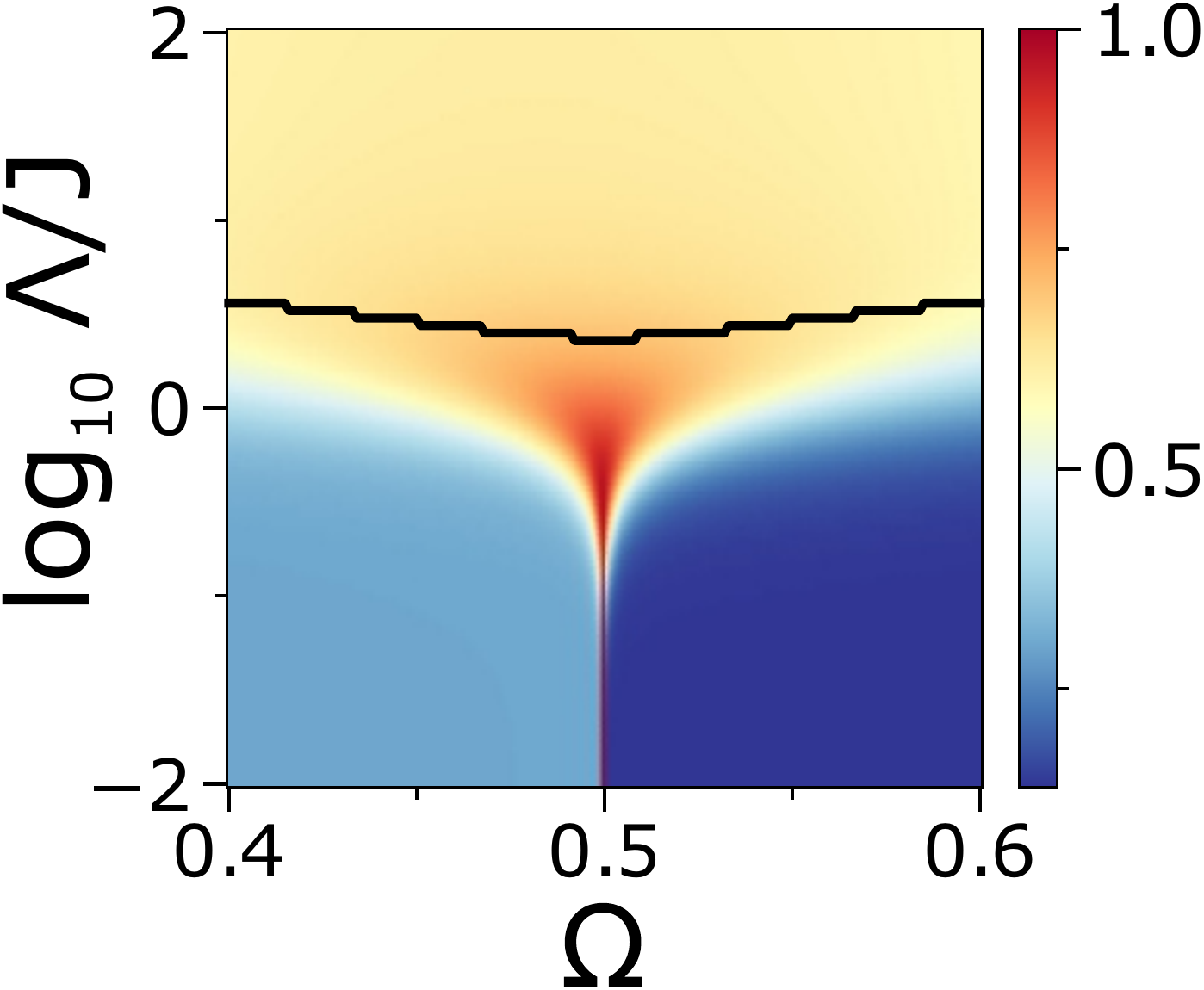}
	\caption{Momentum noise $\sigma_k({\vc{k}=0})$ (in color, normalized to one) plotted for potential barrier $\Lambda$ against \idg{a} on-site interaction $U$ (${\Omega=\frac{1}{2}}$) and \idg{b} flux $\Omega$ (${U/J=1}$). Momentum noise is extracted from time-of-flight image after long expansion. Only ring is expanded, without central condensate. Black line shows the critical point where depletion at the potential barrier is 1\% of the average particle number per site. Above the line the potential barrier site is depleted. Other parameters are ${M=11}$ ring sites and 5 particles.}
	\label{CorrUvsp}
\end{figure}

Although the interferometric pictures look similar (see Fig.\ref{densUtime} for the evolution of the density), different interactions lead to current states that may be very different in nature. For ${U=0}$, the current is in a non-entangled superposition state, whereas for interaction ${U=J}$ in a highly entangled NOON state. 
Below, we shall see how additional  information on the states can be grasped analysing the noise  in the momentum distribution of the ring condensate. Indeed, the  noise for zero momentum depends strongly on the specific entanglement between the clockwise and anti-clockwise flows. 
In the case of an entangled cat state all atoms have together either zero or one momentum quanta. A projective measurement will collapse the wavefunction to either all atoms in the zero or one momentum state. Averaging over many repeated measurements will result in erratic statics of the measurements.
In contrast, in non-entangled single-particle superpositions, each particle has independently either zero or  one momenta quanta. A single projective measurement will result in on average half the atoms having zero and half the atoms having one rotation quantum. Therefore,  fluctuations averaged over many measurements will be low.
We define the noise of the momentum distribution
\begin{equation}
\sigma_k(\vc{k})=\sqrt{\avg{\op{n}(\vc{k})\op{n}(\vc{k})}-\avg{\op{n}(\vc{k})}\avg{\op{n}(\vc{k})}}\;.
\end{equation}
Having in mind a time-of-flight experiment, 
the optimal point to measure the noise is at ${\vc{k}=0}$, as at this point the density is maximal for zero rotation quanta, and zero for one or more rotation quanta.
We plot the noise of the time-of-flight image at ${\vc{k}=0}$ without a central condensate in Fig.\ref{CorrUvsp}. We see that the momentum noise is minimal in a certain parameter regime in Fig.\ref{CorrUvsp}a. This area can provide a guide for  the Gross-Pitaevskii limit of the Bose-Hubbard model. It is given by ${U/J\ll1}$ and ${\Lambda/J>cU/J}$, where $c$ is some constant. As soon the interaction becomes larger than the energy gap induced by the potential barrier, the noise increases. \rev{This indicates the deviation from the Gross-Pitaevskii regime. 
Here, entangled phase winding states of zero and one winding quantum appear. For large interaction, the noise decreases again, however remains higher than in the Gross-Pitaevskii regime.}
As we elaborate further in the Sect.\ref{conclusions},  information about the entanglement is also hidden in the number of particles at the site of the potential barrier\cite{aghamalyan2015coherent}. 
The noise is maximal at the degeneracy point and when barrier and interaction are on the same order. Changing the flux away from the degeneracy point decreases the noise.


\subsection{Gross-Pitaevskii  interferometric measurement}
\begin{figure}
	\includegraphics[width=0.49\textwidth]{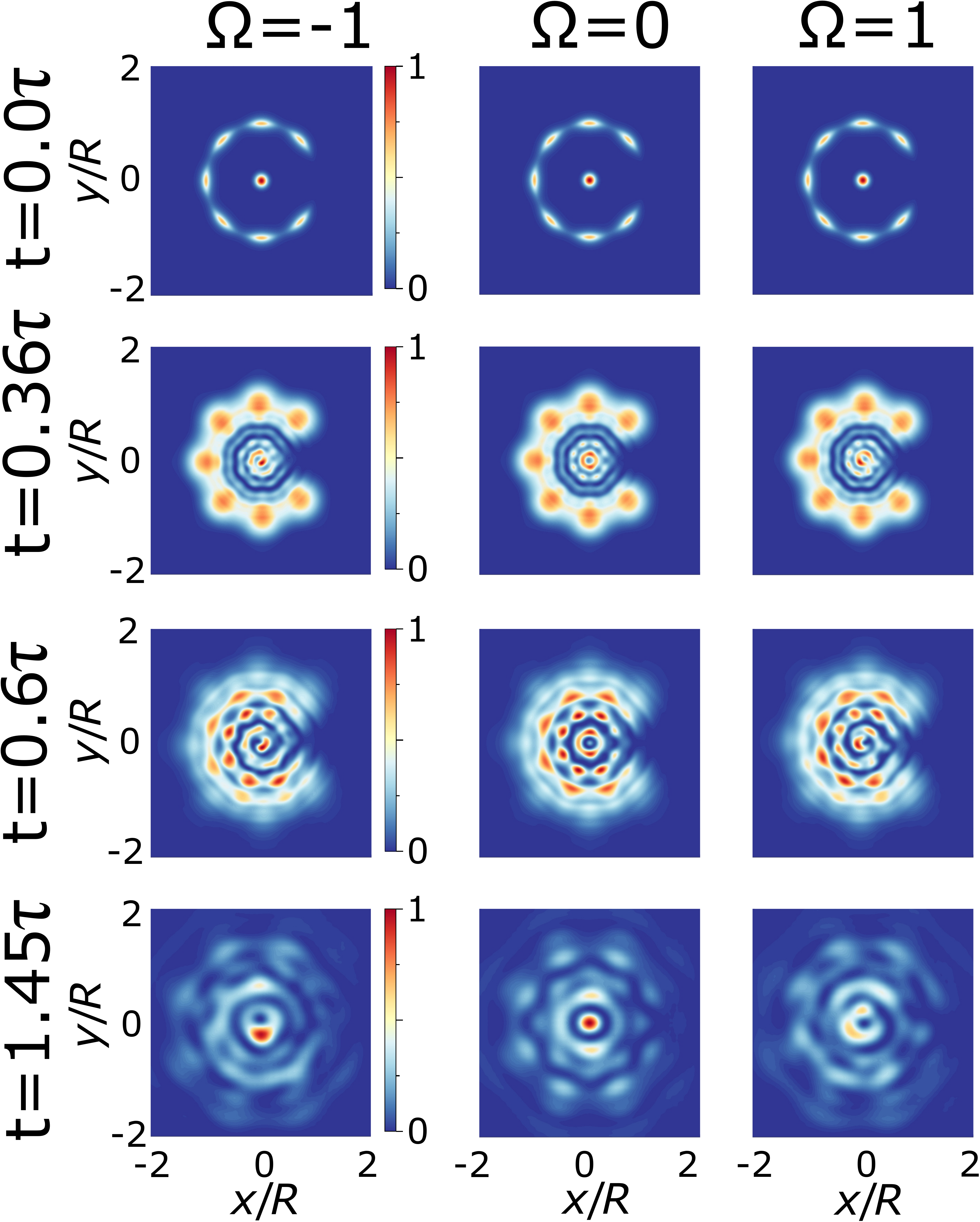}
\caption{Density of the condensate during expansion for different values of $ \Omega $ at different times in units of ${\tau=mR\sigma_r/\hbar}$. Calculated using Gross-Pitaevskii equation.  Parameters are ${M=8}$, $ {\eta_\text{w} = 0.8} $, $ {\eta_\text{l} = 0.5} $, $ {\nnl = 10}$, ${V_0=200\hbar\Omega_0}$, ${\sigma_\text{c}=0.1R}$, $\sigma_\text{r}=0.083R$, ${ \xi_\text{r} = 0.04R }$.}
\label{omgt_largerange}
\end{figure}

\begin{figure*}
	\includegraphics[width=0.9\textwidth]{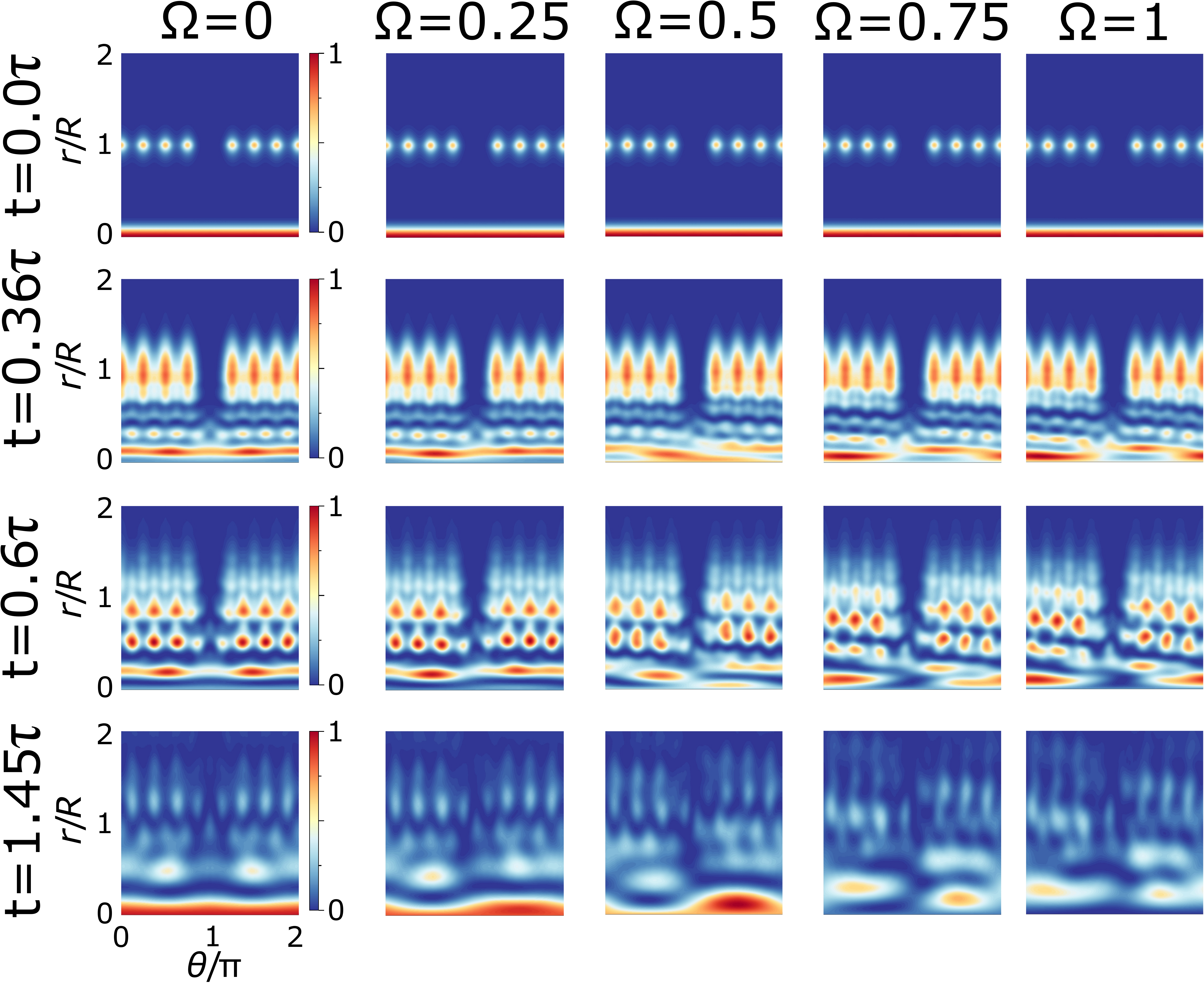}
	\caption{Density of expanding atoms in polar coordinates with radius $r$ and polar angle $\theta$ for increasing $ \Omega $ across the degeneracy point. Other parameters as in Fig.\ref{omgt_largerange}.}
	\label{cmpomg_polar}
\end{figure*}

In Fig.\ref{omgt_largerange}, we plot the ring lattice condensate for different values of flux $\Omega$. For non-zero flux, spirals are clearly observed.  We point out that, in contrast with the BH dynamics presented above, here the atom-atom interaction is kept finite during the expansion.
As in  the Bose-Hubbard dynamics, a discontinuity in the spiral is observed at the weak link site. This discontinuity reflects the phase jump across the weak link of the superposition state (see Figure \ref{cmpomg_polar}). This discontinuity is slightly visible at ${\Omega=0.25}$, indicating that there is a small contribution of one rotation quantum in the condensate. The discontinuity is maximal at ${\Omega=\frac{1}{2}}$. In the polar plot, the spirals are visible as slope in angular direction and are maximal for ${\Omega=1}$. 

\begin{figure}
	\includegraphics[width=0.35\textwidth]{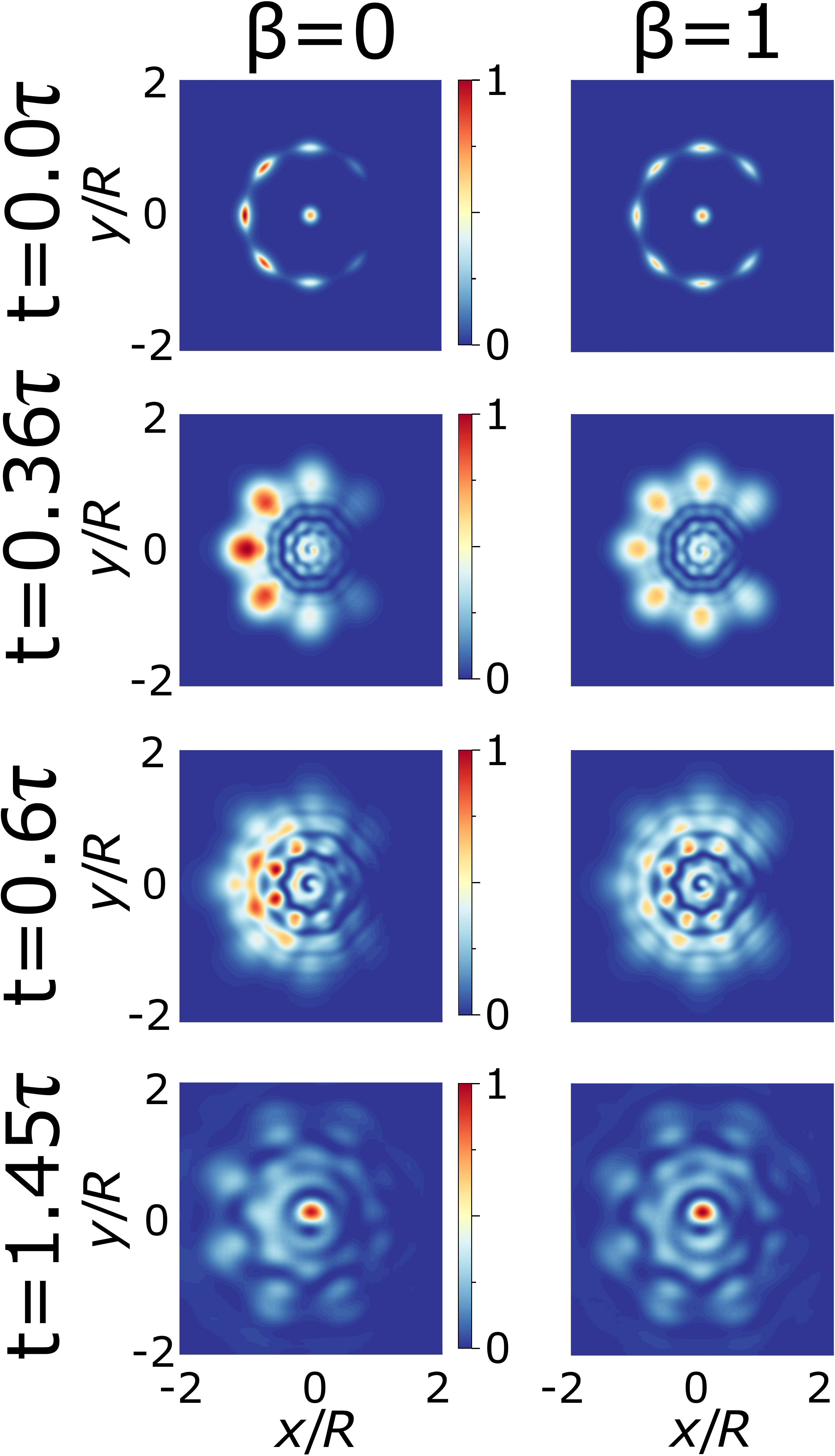}
	\caption{Density of the condensate during expansion for interaction ${\nnl=0}$ (left) and ${\nnl =1}$ (right) at the degeneracy point ${ \Omega = \frac{1}{2}}$. Other parameters as in Fig.\ref{omgt_largerange}.}
	\label{gt}
\end{figure}

In Fig.\ref{gt}, we plot the time evolution of the condensate for two values of the interaction.

\section{Discussion}
\label{conclusions}
In this paper, we have studied the current states in Atomtronic Quantum Interference Devices (AQUIDs) defined as a ring-lattice of condensates interrupted by a weak-link and pierced by an effective magnetic flux. The ring lattice without weak-links has well defined angular momenta states, with degeneracy between two momentum states at ${\Omega=(2n+1)/2}$. With the weak-link, the degeneracy is lifted and the AQUID is in a superposition state of two momenta. The interplay between the weak-link strength and interaction provides a specific crossover\cite{aghamalyan2015coherent}. In particular,  the AQUID can define an effective qubit for strong enough interaction and weak-link strength.
To detect the phase configurations in the AQUID, the ring-shaped condensate is let to interfere with a non-rotating condensate place at the center of the ring (heterodyne phase detection). 
The noise in the time-of-flight image of the expanding ring condensate (without the central cloud) is analyzed.
The dynamics of the combined condensates is studied through Gross-Pitaevskii and Bose-Hubbard dynamics. \rev{As discussed above , the Gross-Pitaevskii regime is investigated as a single-shot expansion; the Bose-Hubbard dynamics, instead, is studied through quantum mechanical expectation values (see Sect.\ref{BHdynamics})}.
\paragraph{Interferometric detection.}
For Gross-Pitaevskii regime with small $U/\Lambda$, the spiral pattern clearly emerges at times ${t<\tau}$ in the density of the expanding condensates, for both vanishing and intermediate interaction--Fig.\ref{omgt_largerange},\ref{gt}.
At increasing times, the spiral is washed out and replaced by a ring-shaped density, which corresponds to the Fourier transform. 
The lattice acts as an azimuthal modulation of the density along the ring. This leaves a lattice-shaped imprint in the time-of-flight image, retaining the distinctive spiral features and phase slips. 
For larger interactions (compared with $\Lambda$), the regime well described by the Bose-Hubbard model, \rev{the read-out is studied through expectation values defined by the average of many condensate expansions. 
In this case, we found that the} spiral evolution is not displayed by density, but by the density-density covariance--Fig.\ref{genFlux}. 
We found  no significant dependence  between the timescales of the expansion and the effective flux. 
\paragraph{Degeneracy point.}
The degeneracy at the effective magnetic flux ${\Omega=(2n+1)/2}$ is reflected by a specific discontinuity  of the spiral pattern at the weak link (a phase jump)--Fig.\ref{corrUtime},\ref{cmpomg_polar}. For the Bose-Hubbard model, we found that for stronger interactions, the discontinuity of the spirals becomes slightly smaller (see Fig.\ref{corrUtime} at ${t=0.6\tau}$). This trend is consistent with the renormalization that the interaction implies on the strength of the local barrier\cite{aghamalyan2015coherent}. 

\rev{While for small interactions, the Gross-Pitaevskii equation produces reasonable superposition states, we find that, with increasing interaction, an instability occurs at the degeneracy point:  In a situation in which we should observe an equal superposition of zero and one phase winding,  we observe that the system tends to acquire  zero (or one) phase winding,  instead. This marks the break-down of the underlying mean-field approximation of the Gross-Pitaevskii equation. In the time evolution, we observe this effect as a vanishing discontinuity of the spiral at the weak link  for increasing interactions (see Appendix Fig.\ref{cmpphase})}.
In fact, the Bose-Hubbard model correctly describes this regime of strong interactions. We found that, because of the two level system effective physics, the noise in the time-of-flight of the ring condensate, {\it without the central cloud},  is particularly pronounced at the degeneracy points--Fig.\ref{CorrUvsp}. This phenomenon would allow to detect the degeneracy point  in the ring condensate, without resorting the heterodyne detection protocol.
\paragraph{Macroscopic quantum coherence.}
With increasing interaction, we can define three regimes of entanglement\cite{nunnenkamp2011superposition}: At the degeneracy point ${\Omega=\frac{1}{2}}$,  for interaction smaller than the energy gap created by the weak link, we observe one-particle superposition states  $\ket{\Psi}\propto(\ket{{l=0}}+\ket{{l=1}})^N$, where $N$ is the number of particles and $l$ is the angular momentum of the atom. This regime is well described by the Gross-Pitaevskii equation. When the interaction and the weak-link energy gap is on the same order, the near-degenerate many body states mix and entangled NOON states are formed $\ket{\Psi}\propto\ket{{l=0}}^N+\ket{{l=1}}^N$. Increasing interaction further will fermionize the system. With interaction, angular momentum of each atom individually is not conserved, however the center of mass angular momentum $K$ of the whole condensate is. Then, the ground state is a superposition of  $\ket{\Psi}\propto\ket{{K=0}}+\ket{{K=N}}$. 

This entanglement can be observed in the noise of the momentum distribution--Fig.\ref{CorrUvsp}.  The ring atoms (without central condensate) are freely expanded until the density distribution of the atoms corresponds to the momentum distribution $\vc{k}$ of the initially prepared state. 
For a non-entangled superposition state, the momentum states have a binomial distribution and the noise at ${\vc{k}=0}$ is minimal (Gross-Pitaevskii regime) and is given by ${\sigma_k^\text{GP}(\vc{k}=0) \propto \sqrt{N}/4}$. For a completely entangled NOON state, it is maximal and given by ${\sigma_k^\text{NOON}(\vc{k}=0) \propto N/4}$. The ratio of the two extrema is $\sqrt{N}$. Thus, with increasing particle number these two types of states are clearer to distinguish. 


For zero on-site interaction, the site at the potential barrier is always depleted at the degeneracy point for any value of potential barrier strength. However, when  the interaction exceeds  a critical value,  particles start occupying the site\cite{aghamalyan2015coherent}. This is plotted as black line in Fig.\ref{CorrUvsp}. For small interaction the critical value has a linear relationship between $U$ and $\Lambda$\cite{aghamalyan2015coherent}. The filling of the potential barrier site indicates the onset of entanglement between different flux quanta. The depletion factor can be measured by a lattice-site resolved absorption measurements.

In summary, the read-out of the AQUID can be done by analysis of the density (GPE) or by  density-density covariance (BH) of  the ring condensate interfering with  a cloud fixing a phase reference (heterodyne detection protocol). The superposition state is reflected by a specific discontinuity of the phase pattern. The entanglement of the state can be studied by looking at the expansion of the ring condensate (without the central cloud) by two witness: First, the degree of depletion at the potential barrier corresponding to the weak-link. For a weak barrier (compared to the interaction), when the barrier site fills up, this indicates the onset of entanglement. Second, the noise in the time-of-flight image. The noise becomes maximal for a highly entangled NOON-state. It is minimal for an one-particle superposition state, and intermediate for center-of mass superpositions.

We believe that our findings  are  well within the current experimental know-how of the field.

\begin{acknowledgments}
{\it Acknowledgments}. The Grenoble LANEF framework (ANR-10-LABX-51-01) is acknowledged for its support with mutualized infrastructure. We thank National Research Foundation Singapore and the Ministry of Education Singapore Academic Research Fund Tier 2 (Grant No. MOE2015-T2-1-101) for support.
\end{acknowledgments}

\begin{appendix}

\section{Additional GPE data}
\begin{figure}[htbp]
	\includegraphics[width=0.49\textwidth]{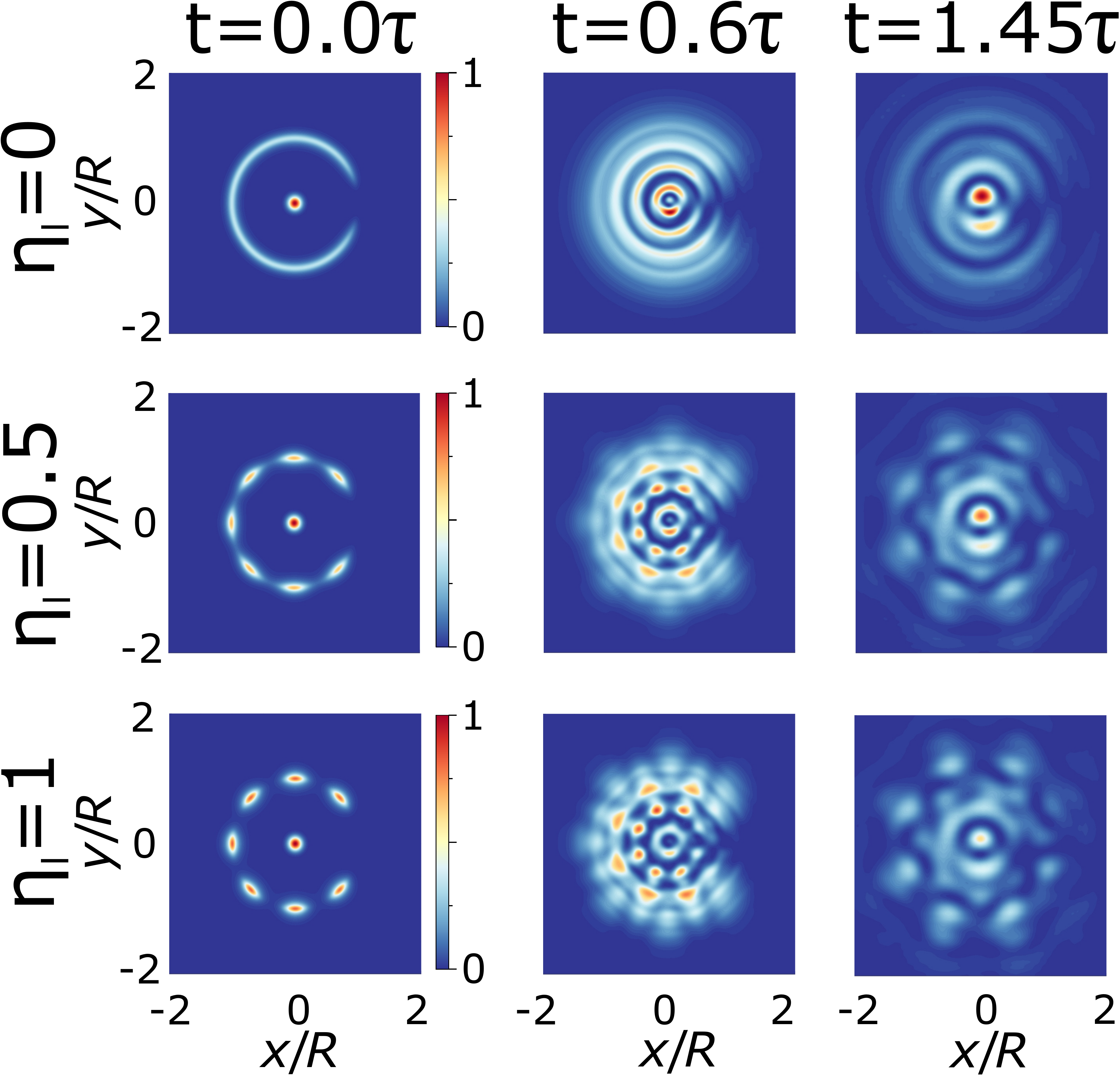}
	\caption{Density of the condensate during expansion for different lattice depths $\eta_\text{l}$ solved with GPE equation. Parameters are ${ \Omega = \frac{1}{2}} $, ${ \nnl = 10}$, all other as in Fig.\ref{omgt_largerange}. }
	\label{ltfig}
\end{figure}
In this section, we provide additional evidence for the dynamic calculated with the Gross-Pitaevskii equation.

In Fig.\ref{ltfig}, we plot the time evolution of ring and central cloud condensate against different lattice depths $\eta_\text{l}$ at the degeneracy point. We see that at intermediate times ${t=0.6\tau}$ the spiral position jump at the weak link site at the center right. This feature is due to the jump of phase at the weak link and is independent of lattice depth. With increasing lattice depth, the time-of-flight image acquires a lattice structure as well.

%

\begin{figure}[htbp]
	\includegraphics[width=0.35\textwidth]{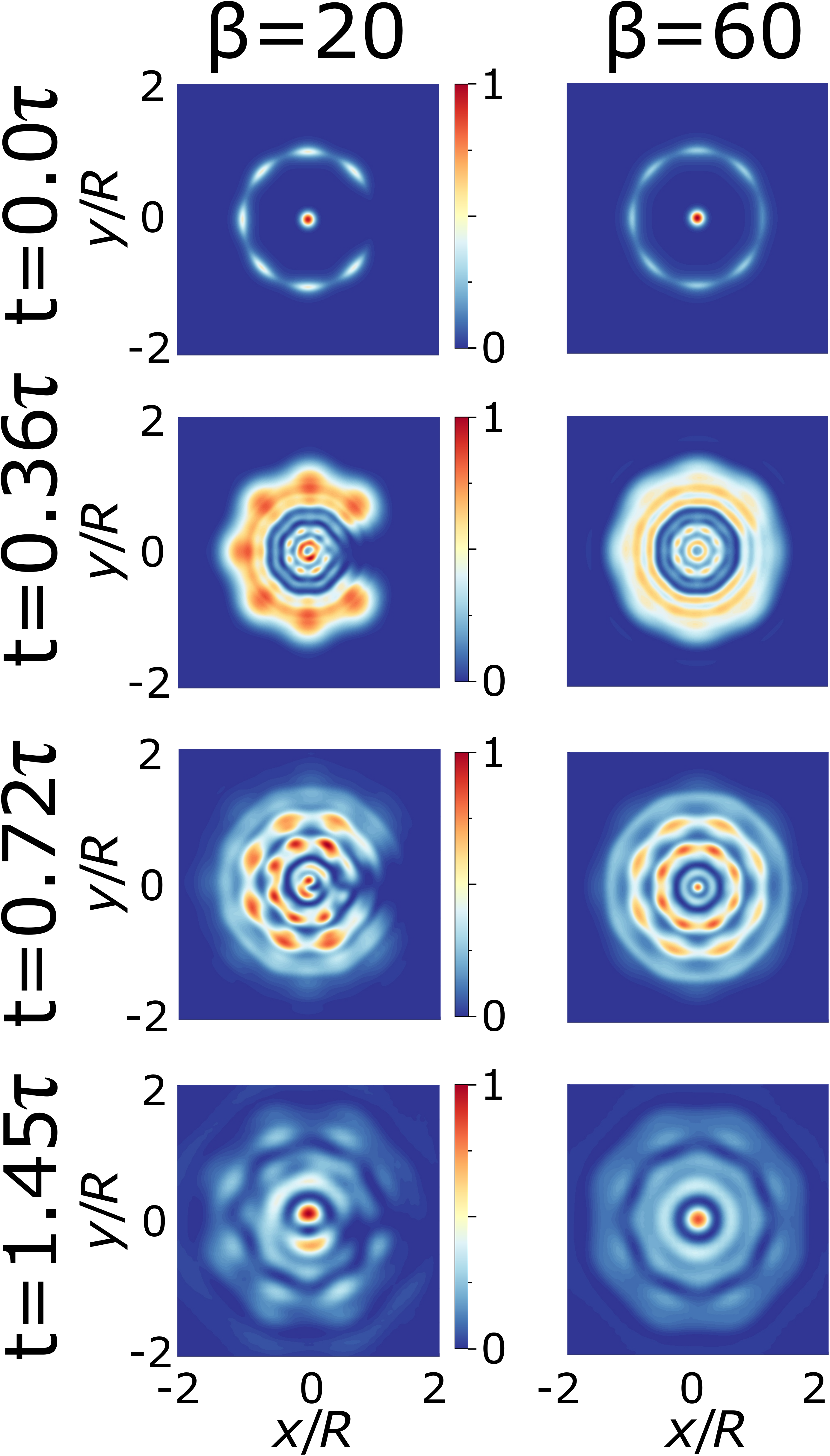}
	\caption{Density of the condensate with Gross-Pitaevskii equation during expansion for $ \eta_w = 0.7 $, $ \nnl = 20  $ (left) and $ \eta_w = 0.3 $, $ \nnl = 60 $ (right). Other parameters as in Fig.\ref{omgt_largerange}.}
	\label{cmpphase}
\end{figure}
In Fig.\ref{cmpphase}, we plot the time evolution of the condensate at the degeneracy point for different values of interaction $\nnl$. Whereas for low interaction, we see initially the characteristic discontinuity, for strong interaction, it is suppressed. The time evolution for strong interaction shows no spiral and thus no phase winding. This is because the GPE mean-field breaks down when the interaction energy is large compared to energy gap induced by the weak link. Depending on the initial conditions, either zero or one phase winding is observed in this case.

\section{Additional Bose-Hubbard data}
\label{AdditionalBH}
\begin{figure}[htbp]
	\centering

	\includegraphics[width=0.45\textwidth]{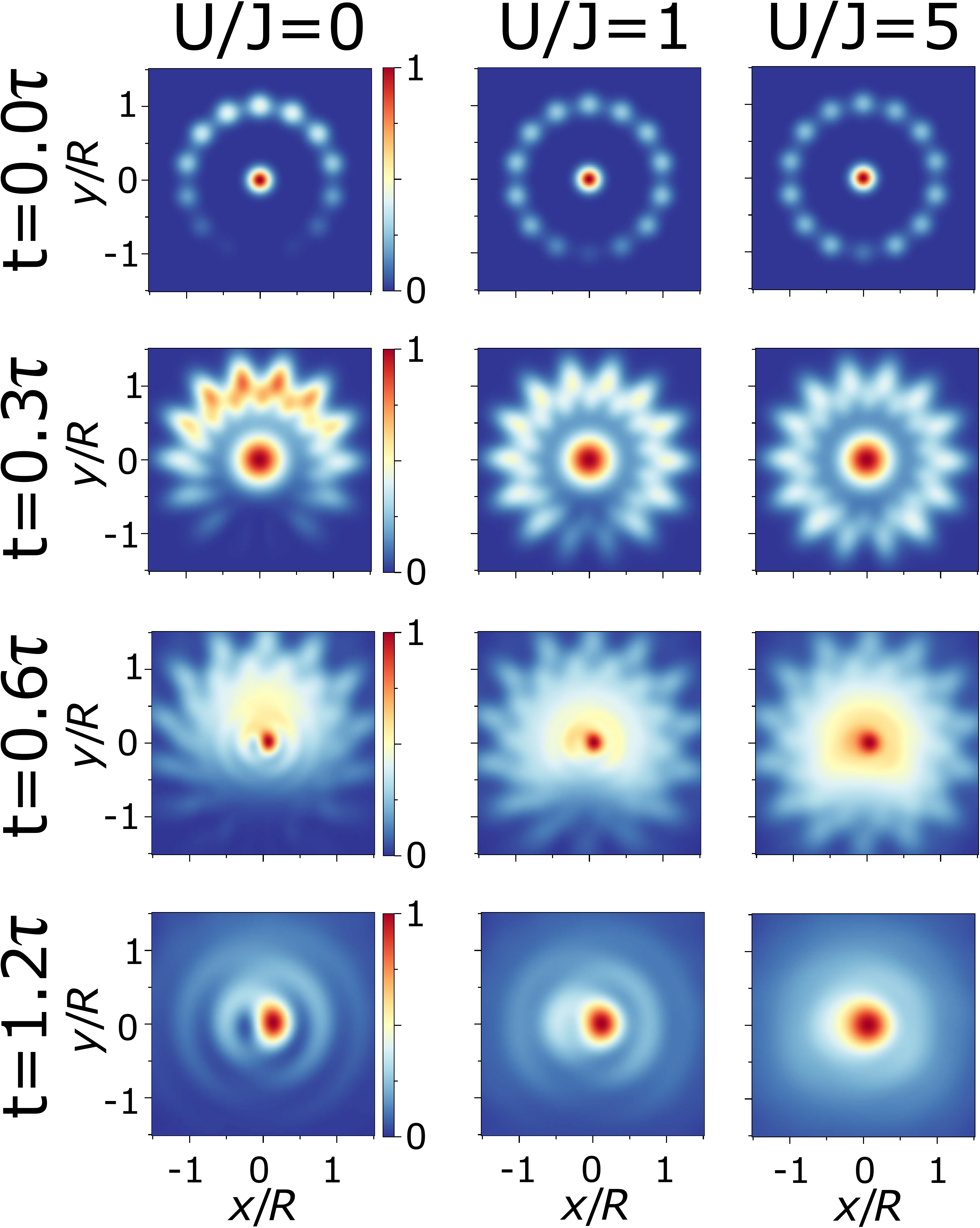}
	\caption{Density of expanded atoms for different interaction and expansion times  for Bose-Hubbard model. Parameters same as in Fig.\ref{corrUtime}. Interaction $U$ in units of $J$.  At intermediate time, we observe some spiral-like structure at the edges. This is not the interference with the central condensate, but a residue of the ring lattice interfering with itself. }
	\label{densUtime}
\end{figure}

In this section, we provide additional data for the Bose-Hubbard simulations. 
First, we plot the density of expanded atoms for different values of interaction at the degeneracy point in Fig.\ref{densUtime}. The corresponding density-density covariance are plotted in the main text in Fig.\ref{corrUtime}. 
The density of expanded atoms at longer times has some characteristic features depending on the interaction. For interaction energy smaller than the potential barrier, the center shows a characteristic bright and dark spot. For stronger interaction, it becomes a single, blurred spot. At the degeneracy point we observe a superposition of counter-flowing current states. Interaction modifies the many-body entanglement (as described in Sect.\ref{conclusions}c), which changes the characteristic time-of-flight pattern. After a long enough free expansion, the atom density assumes the initial momentum distribution. 

\begin{figure*}[htbp]
	\centering

	\includegraphics[width=0.9\textwidth]{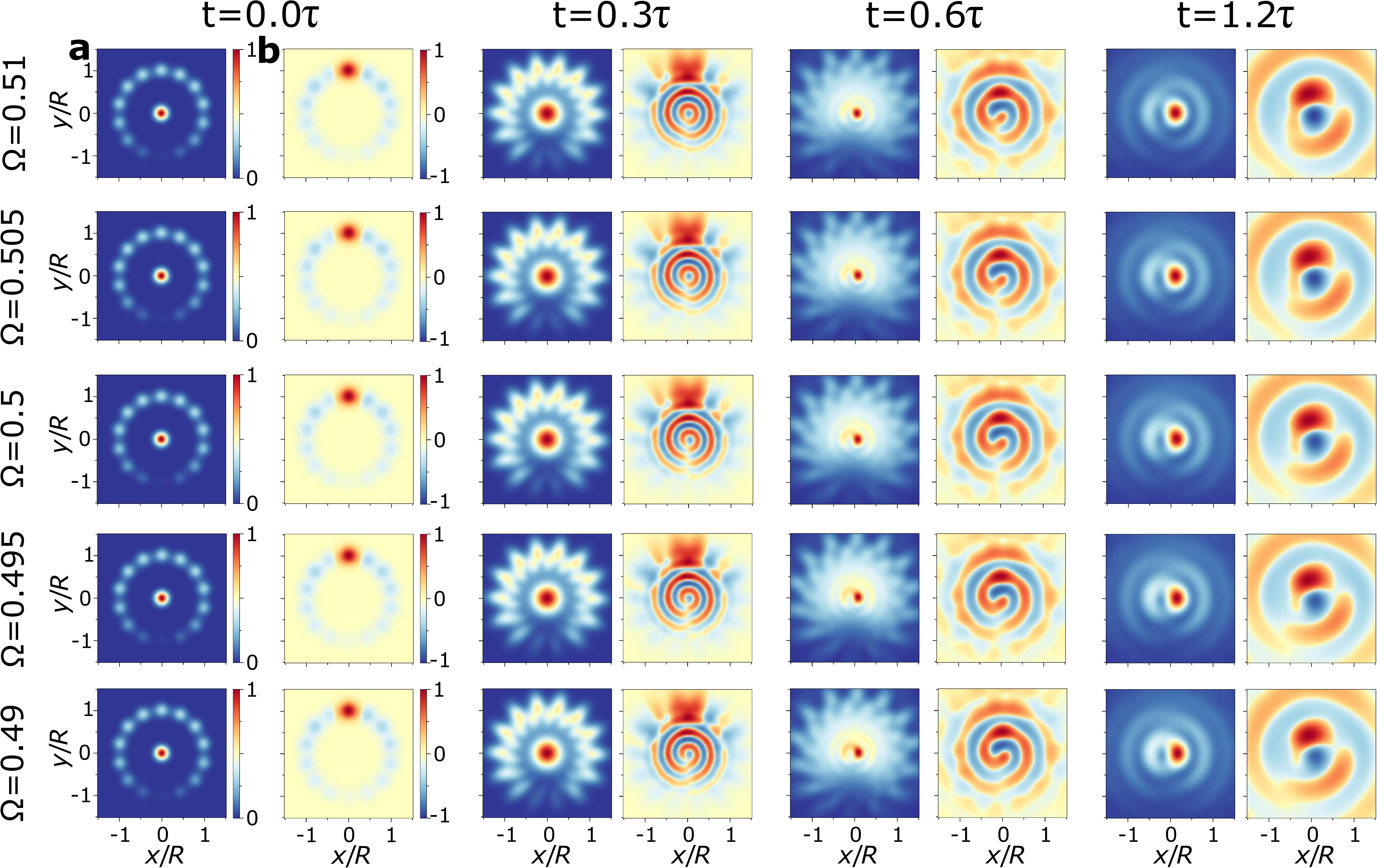}
	\caption{Density \idg{a} and \idg{b} root of density-density covariance ${\sigma(\vc{r},{\vc{r'}=\{0,R/2\}})}$ of expanding atoms at different times for different values of flux across the degeneracy point for the Bose-Hubbard model. Parameters are: 7 particles, ${M=14}$ ring sites, ring radius $R$, width of central and ring cloud ${\sigma_r=2 R/M}$, ${\Lambda=J}$, ${U/J=0.5}$ and 25\% of atoms in central condensate. Barrier at ${x=0}$, ${y=-R}$.}
	\label{densFtime}
\end{figure*}

Finally, we look at the transition across the degeneracy point for different values of $\Omega$ close to the degeneracy point. In Fig.\ref{densFtime}, we plot the density and the density-density covariance. At intermediate times, this graphs shows how the discontinuity of the spirals develops close to the degeneracy point ${\Omega=0.5}$.

\section{Density-density covariance}
\label{AppDens}
We define the density operator $\op{n}(\vc{r},t)=\op{\psi}^\dagger (\vc{r},t)\op{\psi}(\vc{r},t)$, represented by the Wannier functions  $\op{\psi}(\vc{r})=\sum_n w_n(\vc{r})\an{a}{n}$. This defines a proper density if the operator at different positions commute.
The commutator of the density operator is
\begin{align*}
&{}[\op{n}(\vc{r}),\op{n}(\vc{r'})]=\\
&{}\sum_{n,m}(\alpha (\vc{r},\vc{r'}) w_n^*(\vc{r'})w_m(\vc{r})- \alpha^* (\vc{r},\vc{r'})w_n^*(\vc{r})w_m(\vc{r'})) \cn{a}{n} \an{a}{m} \; ,
\end{align*}
with $\alpha (\vc{r},\vc{r'})=\sum_nw_n(\vc{r})w_n^*(\vc{r'})$. The density operators will commute in general only when $\alpha (\vc{r},\vc{r'})=\delta(\vc{r}-\vc{r'})$, which means that the Wannier functions form a complete basis. 
Then, the density-density correlator is
\begin{equation} \op{n}(\vc{r})\op{n}(\vc{r'})=\sum_{i,j,n,m}w_i(\vc{r})^*w_j(\vc{r})w_n(\vc{r'})^*w_m(\vc{r'})\cn{a}{i}\an{a}{j}\cn{a}{n}\an{a}{m}\;.
\end{equation}
We now restrict ourselves to a limited number $N$ of Wannier functions, which we fill with particles (denoted as $\an{a}{j}$), and the rest $M$  is empty (denoted as $\an{b}{x}$). Then we get
\begin{align*} \op{n}(\vc{r})\op{n}(\vc{r'})=&\sum_{i,j,n,m}^Nw_i(\vc{r})^*w_j(\vc{r})w_n(\vc{r'})^*w_m(\vc{r'})\cn{a}{i}\an{a}{j}\cn{a}{n}\an{a}{m}+\\
&\sum_{i,j}^N\sum_{x}^Mw_i(\vc{r})^*w_x(\vc{r})w_x(\vc{r'})^*w_j(\vc{r'})\cn{a}{i}\an{b}{x}\cn{b}{x}\an{a}{j}
\;.
\end{align*}
If we now only calculate the $N$ Wannier functions (first term) and ignore the second term, the density operator at different positions do not commute. As a result, the density-density correlator acquires a complex part.  However, if we assume that the particle number $P$ is very large, then the first term scales as $P^2$, while the second as $P$ (as $\an{b}{x}\cn{b}{x}=1$). 
Thus, neglecting the Wannier functions corresponding to empty sites will yield a smaller error with increasing particle number. We confirmed numerically that with increasing particle number the relative size of the imaginary part becomes negligible (calculated for up to 14000 particles by truncating the Hilbert space). We find that the shape of the density-density covariance and the density does not change significantly for increasing particle number.


\section{Time-of-flight analytics without interaction}
When the condensate is released during time-of-flight analysis, the spatial distribution approaches the momentum distribution of the initial condensate wavefunction. In this section, we show this relationship.
The free-particle Schrodinger equation (we set ${\hbar=1}$, ${m=1}$)
\begin{equation}
i \partial_t \ket{\psi} = \frac{1}{2}\hat{p}^2 \ket{\psi}
\end{equation}

has the following propagator:

\begin{equation}
\begin{split}
K(x, t; x') &= \int_p \braket{x}{p} \braket{p}{x'} e^{-\frac{1}{2} i p^2 t} dp \\
&= \int_p e^{i \left(p \cdot x - \frac{1}{2} i p^2 t\right)} \braket{p}{x'} dp
\end{split}
\end{equation}

\begin{equation}
\braket{x}{\psi\left(t\right)} = \int_p e^{i t \left(\frac{1}{t} p \cdot x - \frac{1}{2} p^2\right)} \braket{p}{\psi_0} dp
\end{equation}

In the limit where $ t \rightarrow \infty $, we can find an analytic solution using the method of steepest descent. Identifying the unique saddle point

\begin{equation}
\nabla_p \left(\frac{1}{t} p \cdot x - \frac{1}{2} p^2\right) = 0 \implies p = \frac{x}{t}
\end{equation}

we arrive at the solution

\begin{equation}
\left|\braket{x}{\psi\left(t\right)}\right| = \frac{1}{\mathcal{N}}\left|\braket{p = \frac{x}{t}}{\psi_0}\right|
\end{equation}

where $ \mathcal{N} $ is a normalisation factor.

\section{Perturbation Theory Analysis close to the degeneracy point}
\label{appendix_perurbation}
An approximate solution to the ground state energies of the non-interacting system can be found using degenerate perturbation theory. We start by idealising the trap potential of Eq.\ref{potential_equation}. We assume that the potential highly confines the condensate to a quasi one-dimensional system and the weak link is a Delta-function.

\begin{equation}
V = -V_0\delta\left(r - R\right)\left(1 - \eta\delta\left(\theta\right)\right)
\end{equation}

We consider only the azimuthal part of the Hamiltonian and gauge away the constant term $ V_0 $. We can then write the weak link as a perturbation to the rotating-frame Hamiltonian:

\begin{equation}
\begin{split}
H &= H_0 + H' \\
H_0 &= -\frac{1}{2}\partial_\theta^2 + i\Omega\partial_\theta \\
H' &= V_0\eta\delta\left(\theta\right)
\end{split}
\end{equation}

When $ \Omega = \frac{1}{2} $, the degenerate ground states of the exactly solvable Hamiltonian $ H_0 $ with energy $ E_g = 0 $ are

\begin{equation}
\ket{0} = \frac{1}{\sqrt{2\pi}},\
\ket{1} = \frac{1}{\sqrt{2\pi}} e^{i\theta}
\end{equation}

We can then write the perturbation $ H' $ as a matrix in $ \ket{0} $ and $ \ket{1} $:

\begin{equation}
H' = V_0\eta\begin{bmatrix} 1 & 1 \\ 1 & 1 \end{bmatrix}
\end{equation}

This matrix has eigenvalues

\begin{equation}
E_0' = 0,\ 
E_1' = 2V_0\eta
\end{equation}

and corresponding eigenvectors

\begin{equation}
\ket{0'} = \frac{1}{\sqrt{2}}\begin{bmatrix} 1 \\ -1 \end{bmatrix},\ 
\ket{1'} = \frac{1}{\sqrt{2}}\begin{bmatrix} 1 \\ 1 \end{bmatrix}
\end{equation}

resulting in the spectrum

\begin{equation}
E_0 = E_g + E_0' = 0,\ 
E_1 = E_g + E_1' = 2 V_0 \eta
\end{equation}

When the weak link is not a delta-function, but has a finite width,  a small correction appears in the off-diagonal elements of $ H' $. For a narrow Gaussian weak link with width ${\xi\ll2\pi}$ the perturbation becomes
\begin{equation*}
H'=V_0\eta\frac{1}{\sqrt{2\pi\xi^2}}\expU{-\frac{\theta^2}{2\xi^2}}\;,
\end{equation*}
and we get the matrix elements
\begin{equation}
H' = V_0\eta\begin{bmatrix} 1 &\expU{-\frac{\xi^2}{2}} \\ \expU{-\frac{\xi^2}{2}} & 1 \end{bmatrix}
\end{equation}

This matrix has the same eigenvectors, but with eigenvalues

\begin{equation}
E_0' = V_0 \eta \left(1-\expU{-\frac{\xi^2}{2}}\right),\ 
E_1' = V_0 \eta \left(1+\expU{-\frac{\xi^2}{2}}\right)
\end{equation}
In the limit $\xi\rightarrow0$, the result corresponds to the one with the delta function. A broader weak link reduces the energy gap induced by the weak link.

\end{appendix}

\bibliography{library}

\end{document}